\documentclass{aa}
\usepackage{graphicx}
\usepackage{natbib}
\bibpunct{(}{)}{;}{a}{}{,}
\begin{document}

\title{$^{12}$CO(J=2$\to$1) and CO(J=3$\to$2) observations of Virgo 
 Cluster spiral galaxies with the KOSMA telescope: global properties}

\titlerunning{$^{12}$CO observations of Virgo Cluster spiral galaxies}

\author{H. Hafok \inst{1,2} \and J. Stutzki \inst{1}}

\institute{I. Physikalisches Institut, Universit\"at zu K\"oln, 
           Z\"ulpicher Strasse 77, 50937 Cologne, Germany
      \and Radioastronomisches Institut, Universit\"at Bonn, 
           Auf dem H\"ugel 71, 53121 Bonn, Germany}

\offprints{J. Stutzki}

\date{Received 02-JUL-02 / Accepted 12-NOV-02 }

\abstract{We present $^{12}$CO\,(J=2$\to$1) and CO\,(J=3$\to$2)
  observations of quiescent Virgo Cluster spiral galaxies with the
  KOSMA 3m submm telescope. The beam sizes of 80\arcsec{ }at
  345\,\rm{GHz} and 120\arcsec{ }at 230\,\rm{GHz} are well suited for
  the investigation of global properties of Virgo Cluster galaxies.
  The observed sample was selected based on previous
  $^{12}$CO\,(J=1$\to$0) detections by \citet{Stark86}, performed with
  the AT\&T Bell Laboratory 7m telescope (beam size $\sim$
  100\arcsec). We were able to detect 18 spiral galaxies in
  $^{12}$CO\,(2$\to$1) and 16 in $^{12}$CO\,(3$\to$2).  Beam matched
  observations of the lowest three $^{12}$CO transitions allow us to
  compare our results with previous high spatial resolution studies of
  (moderate) starburst galaxies and galactic core regions. We discuss
  the global excitation conditions of the ISM in these quiescent
  spiral galaxies. The resulting CO (3--2)/(1--0) integrated line
  ratios vary over a relatively narrow range of values from 0.35 to
  0.14 (on a $\rm{K km/s}$--scale) with increasing CO (2--1)/(1--0)
  ratio (from 0.5 to 1.1). The line ratios between the three lowest
  rotational transitions of CO cannot be fitted by any radiative
  transfer model with a single source component. A two-component
  model, assuming a warm, dense nuclear and a cold, less dense disc
  component allows us to fit the observed line ratios for most of the
  galaxies individually by selecting suitable parameters. The
  two-component model, however, fails to explain the observed
  correlation of the line ratios. This is due to a variation of the
  relative filling factor of the warm gas alone, assuming a typical
  set of parameters for the two components common for all galaxies.
  }

\maketitle

\section{Introduction}

We present $^{12}$CO\,(J=2$\to$1) and $^{12}$CO\,(J=3$\to$2)
observations of 20 spiral galaxies in the Virgo Cluster. Several
surveys in $^{12}$CO have been carried out to study the properties of
the cold ISM in external galaxies with large telescopes like the IRAM
30m or FCRAO 15m e.g. \citet{Young95}, \citet{braine92}. Most of these
are limited to the lowest rotational transitions of $^{12}$CO
(J=1$\to$0) and (J=2$\to$1) which have upper state energies above
ground of 5.3\,\rm{K} and 17\,\rm{K}.  Improvements in receiver and
telescope technology during the last years have made it feasible to
study higher CO-transitions in the submillimeter domain in order to
trace the warmer and denser molecular gas. With an energy above ground
state of 33\,\rm{K} the $^{12}$CO\,(J=3$\to$2) transition is in the
focus of the present investigation. This survey increases the
extragalactic sample of higher CO transitions like $^{12}$CO\,(3--2),
so far limited to a few mapping projects, e.g.  by
\citet{Wielebinski99} and to surveys of nearby galaxies, e.g. by
\citet{Devereux94}, \citet{Mauersberger99}.  Because of the
observational difficulties concerning weather conditions and telescope
efficiencies, in most cases the mid--J CO surveys are biased towards
nearby spirals with strong CO-lines and galaxies with (moderate)
starbursts. These surveys were exclusively made with large ($>$10\,m)
telescopes like the JCMT, HHT or CSO. The main beam of these
telescopes is much smaller than the CO emitting region in the observed
galaxies. Thus accurate global fluxes are difficult to determine, even
for the few galaxies studied at the distance of the Virgo Cluster with
CO-diameters of around 80\arcsec. With the survey observations
restricted to only a few points within the sample galaxy the question
arises how characteristic the deduced line properties are
for the observed galaxy.

This is the first study of global properties of $^{12}$CO\,(2--1) and
$^{12}$CO\,(3--2) in quiescent spiral galaxies in the Virgo Cluster.
At the adopted distance of 15\,Mpc \citep{Stark86} the beam sizes are
equal to 7-10\,kpc and cover at least the core region and inner disc.
Most of the CO emission is thus contained in a single pointing of the
telescope beam and gives us the opportunity to deduce the global
properties of the bulk of the molecular ISM in the observed
galaxies.

\begin{table*}[ht]
\centering{
\caption{\footnotesize Galaxy parameters of the observed sample of spiral 
          galaxies in the Virgo Cluster}
\begin{tabular}{llrrrrrr}
\hline
source name & RC3 type  && inclination & R.A. & DEC & v$_{\rm lsr}$& $Dist_{\rm M87}$ \\
                       &&& (\rm{deg}) & (B1950.0)  & (B1950.0)&(\rm{km/s})& (\rm{deg})\\
\hline
NGC 4192  &  SAB(s)ab  & 2 & 74 & 12:11:15.5  & 15:10:23 & -135 & 4.8\\
NGC 4212  &  SAc       & 5 & 47 & 12:13:06.6  & 14:10:46 &  -83 & 4.0\\
NGC 4237  &  SAB(rs)bc & 4 & 47 & 12:13:06.6  & 14:10:46 & 1123 & 8.0\\
NGC 4254  &  SA(s)c    & 5 & 28 & 12:16:16.9  & 14:41:46 & 2405 & 3.3\\
NGC 4298  &  SA(rs)c   & 5 & 67 & 12:19:00.4  & 14:53:03 & 1136 & 3.2\\
NGC 4302  &  Sc:sp     & 5 & 90 & 12:19:10.2  & 14:52:43 & 1150 & 3.1\\
NGC 4303  &  SAB(rs)bc & 4 & 25 & 12:19:21.4  & 14:44:58 & 1568 & 8.2\\
NGC 4321  &  SAB(s)bc  & 4 & 28 & 12:20:23.2  & 16:06:00 & 1575 & 3.9\\
NGC 4402  &  Sb        & 3 & 75 & 12:23:35.8  & 13:23:22 &  234 & 1.4\\
NGC 4419  &  SB(s)a    & 1 & 67 & 12:24:24.6  & 15:19:24 & -209 & 2.8\\
NGC 4501  &  SA(rs)b   & 3 & 58 & 12:29:28.1  & 14:41:50 & 2284 & 2.1\\
NGC 4527  &  SAB(s)bc  & 4 & 72 & 12:31:35.5  & 14:41:50 & 2284 & 9.8\\
NGC 4535  &  SAB(s)c   & 5 & 43 & 12:29:28.1  & 14:41:50 & 2284 & 4.3\\
NGC 4536  &  SAB(rs)bc & 4 & 67 & 12:31:47.9  & 08:28:25 & 1962 &10.2\\
NGC 4567  &  SA(rs)bc  & 4 & 46 & 12:34:01.1  & 11:32:01 & 2277 & 1.8\\
NGC 4568  &  SA(rs)bc  & 4 & 64 & 12:34:03.0  & 11:30:45 & 2255 & 1.8\\
NGC 4569  &  SAB(rs)ab & 2 & 63 & 12:34:18.7  & 13:26:18 & -236 & 1.7\\
NGC 4647  &  SAB(rs)c  & 5 & 36 & 12:41:01.1  & 11:51:21 & 1422 & 3.2\\
NGC 4654  &  SAB(rs)cd & 6 & 52 & 12:41:25.7  & 13:23:58 & 1039 & 3.3\\
NGC 4689  &  SA(rs)bc  & 4 & 30 & 12:45:15.2  & 14:02:07 & 1620 & 4.3\\
\hline
\end{tabular}
}
\end{table*}

\section{The observed sample}
Our sample of spiral galaxies is based on the Virgo Cluster
$^{12}$CO\,(1--0) survey by \citet{Stark86}, performed with the AT\&T
Bell Laboratory 7m telescope (beam size 100\arcsec) and the
$^{12}$CO\,(1--0) survey by \citet{Ken88}, performed with the FCRAO
telescope (beam size 45\arcsec). Both samples are blue
magnitude-limited (B$_{\rm T}^{\rm 0}$=12) (\citet{deVau76}, \citet{Sandage81})
and consist of spiral galaxies (morphological type Sa or later). From
these two surveys the 20 strongest $^{12}$CO\,(1--0) sources, listed in
Table 1, have been selected. We adopt a distance of 15\,Mpc to the
Virgo Cluster \citep{Stark86}.  For three of the galaxies, the
CO\,(1--0) line intensities quoted by \citet{Stark86} and \citet{Ken88}
do not agree, so that we leave the entries blank in Table 2.

For the selected galaxies, previous papers claim a distance variation
of at most $\pm$20\%, so that we can directly compare the observed
CO-fluxes of the different sample galaxies.

\section{Observations and data reduction}
The observations described in this paper have been carried out with
the K\"olner Observatorium f\"ur Submillimeter Astronomie (KOSMA) 3m
submm telescope \citep{Degiacomi95} at 230\,\rm{GHz} during autumn
1999 and at 345\,\rm{GHz} in January 2000. The main beam efficiency
has been determined by observing planets with a continuum backend. It
was found to be 68\% at 230\,\rm{GHz} and 345\,\rm{GHz}. Broadband
Acousto Optical Spectrometers with bandwidths of 1\,\rm{GHz},
corresponding to 1300\,\rm{km/s} at 230\,\rm{GHz} and 800\,\rm{km/s}
at 345\,\rm{GHz}, have been used for the observations.  This is
suitable even for edge-on galaxies with very broad lines of more than
500\,\rm{km/s}. Typical peak antenna temperatures of the observed
galaxies in the Virgo Cluster are between 4 and 30\,\rm{mK} for
$^{12}$CO\,(2--1) (beam size: 110\arcsec) and $^{12}$CO\,(3--2)
(beam size: 80\arcsec).

Because of the weakness of the observed sources, integration times of
typically 3--6\,h were necessary to yield a low noise rms of
1.5\,--3\,mK for $\Delta$v\,=\,10\,\rm{km/s} and
T$_{\rm sys}$\,=\,150\,--\,200\,\rm{K}\,(DSB).

All measurements were performed in a Dual Beam Switch (i.e.\ chop and
nod) technique, with chopper throws of 6\arcmin{ }in azimuth. The DBS
technique yields spectra with flat baselines over the whole bandpass.
For the data reduction it was sufficient to remove a linear baseline.
Due to the fact that some galaxies of the sample are seen edge-on with
very broad lines, the error of the line integral depends on the number
of the baseline channels outside the line. Therefore the error was
determined using
\begin{equation}
\Delta\,I= T_{\rm rms}\Delta \rm{v} \sqrt{N_{\rm line}\left(1+\frac{N_{\rm line}}{N_{\rm baseline}}\right)}
\end{equation}
with N$_{\rm line}$ the number of channels in the line, N$_{\rm baseline}$ the
number of baseline channels, $\Delta \rm{v}$ the channel width, and
T$_{\rm rms}$ the baseline rms per velocity channel, as already
independently used by \citet{Lavezzi99}.

\begin{figure*}[!h]
  \centering 
  \includegraphics[width=15cm]{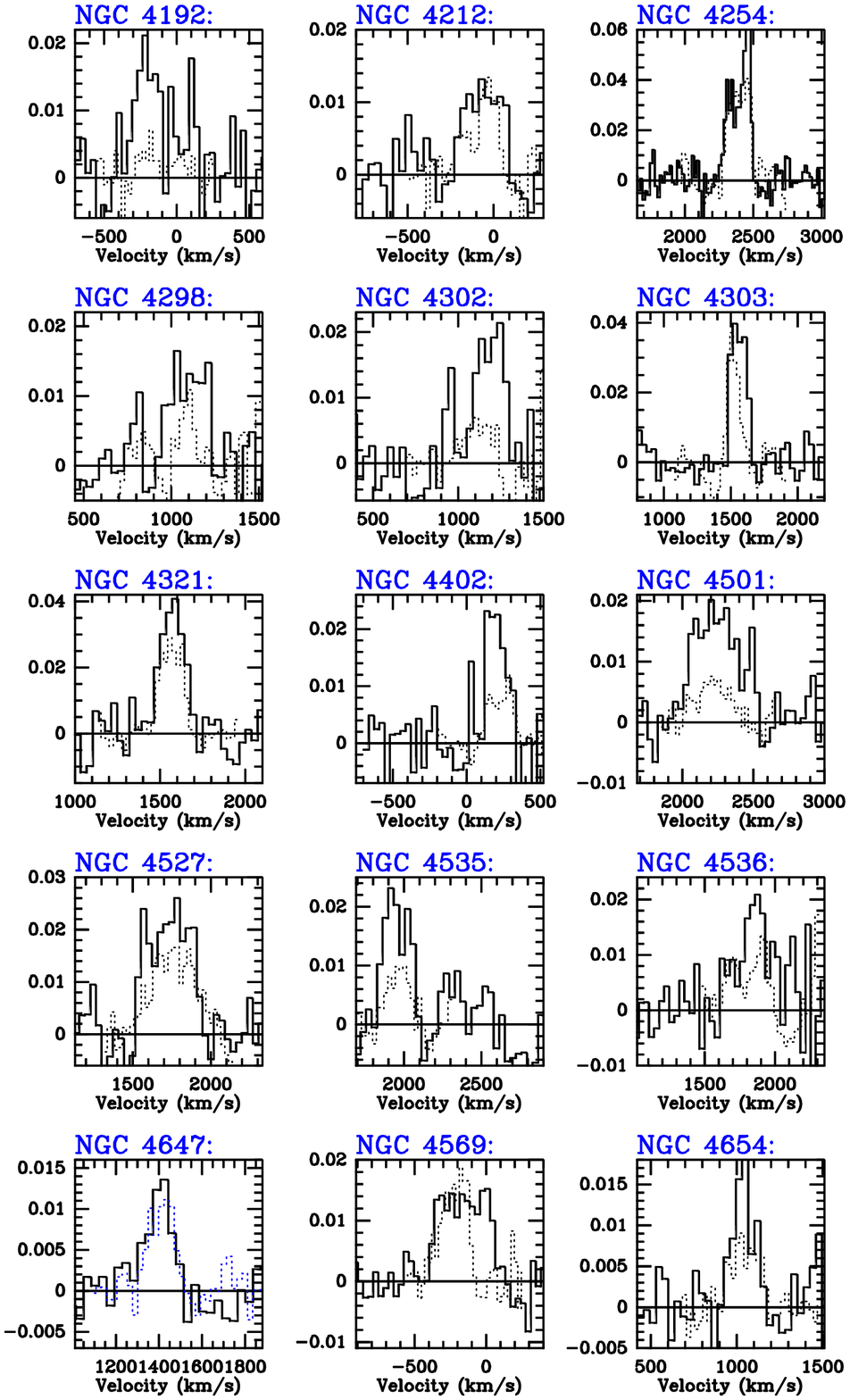}
  \caption{Overview of the sample of Virgo Galaxies detected with the KOSMA 
  telescope (solid:$^{12}$CO\,(2--1), dotted: $^{12}$CO\,(3--2)) at a 
  main beam temperature scale ($\eta$$_{\rm MB}$=0.7 for both transitions).}
\end{figure*}

\begin{table*}[ht]
\centering{
\caption{\footnotesize Observed line parameters of the detected sample 
 galaxies.}
\begin{tabular}{rlllllllllllll}
\hline
Name & I(CO) & $\Delta$I& I(CO) & $\Delta$I & I(CO) & $\Delta$I&$\Theta$ & R$_{\rm 2,1}$&$\Delta R$ & R$_{\rm 3,2}$ & $\Delta R$ & R$_{\rm 3,1}$ & $\Delta R$\\   
NGC    &  (1--0)  &  $1\sigma$    & (2--1) &$1\sigma$&   (3--2)&$1\sigma$&(\arcsec)&&$1\sigma$&& $1\sigma$&& $1\sigma$\\
(1)&(2)&(3)&(4)&(5)&(6)&(7)&(8)&(9)&(10)&(11)&(12)&(13)&(14)\\ 
\hline
4192 & 4.13  & 0.36 & 5.59  & 1.30 &   0.84 & 0.29 & 74.5 & 1.73 & 0.42 & 0.09 & 0.04 & 0.16 & 0.06\\ 
4212 & 4.66  & 0.21 & 3.10  & 0.73 &   2.11 & 0.34 & 48.0 & 0.90 & 0.22 & 0.35 & 0.10 & 0.31 & 0.05\\  
4254 & 13.69 & 0.54 & 8.42  & 0.57 &   3.92 & 0.75 & 57.6 & 0.82 & 0.06 & 0.26 & 0.05 & 0.21 & 0.04\\  
4298 & ----- & ---- & 3.38  & 0.59 &   0.75 & 0.34 & 84.0 & ---- & ---- & 0.14 & 0.07 & ---- & ----\\   
4302 & ----- & ---- & 3.84  & 0.51 &   1.01 & 0.53 & 68.4 & ---- & ---- & 0.15 & 0.08 & ---- & ----\\   
4303 & 10.16 & 0.30 & 5.58  & 0.40 &   3.45 & 0.51 & 75.6 & 0.70 & 0.05 & 0.37 & 0.06 & 0.26 & 0.04\\   
4321 & 12.59 & 0.38 & 6.56  & 0.70 &   4.09 & 0.28 & 60.0 & 0.69 & 0.12 & 0.35 & 0.14 & 0.24 & 0.02\\   
4402 & 4.7   & 0.28 & 4.15  & 0.59 &   1.61 & 0.17 & 90.0 & 1.10 & 0.17 & 0.25 & 0.07 & 0.27 & 0.03\\   
4501 & 13.31 & 0.38 & 6.89  & 0.96 &   1.86 & 0.26 & 76.8 & 0.66 & 0.09 & 0.16 & 0.03 & 0.11 & 0.02\\   
4527 & 17.77 & 0.41 & 7.69  & 0.75 &   5.22 & 0.52 & 91.2 & 0.53 & 0.05 & 0.44 & 0.06 & 0.24 & 0.02\\   
4535 & 5.26  & 0.30 & 4.21  & 0.51 &   1.56 & 0.30 & 62.4 & 1.05 & 0.14 & 0.21 & 0.05 & 0.22 & 0.04\\   
4536 & ----- & ---- & 4.93  & 1.06 &   2.62 & 0.20 & 50.4 & ---- & ---- & 0.28 & 0.06 & ---- & ----\\   
4567 & 5.98  & 0.59 & 5.19  & 0.70 &   1.17 & 0.30 & 48.0 & 1.17 & 0.19 & 0.12 & 0.03 & 0.14 & 0.04\\   
4568 & 8.72  & 0.50 & 5.18  & 0.66 &   4.00 & 0.33 & 68.4 & 0.77 & 0.11 & 0.45 & 0.07 & 0.35 & 0.03\\   
4569 & 11.84 & 0.32 & 5.81  & 0.69 &   3.55 & 0.68 & 68.4 & 0.64 & 0.08 & 0.35 & 0.08 & 0.23 & 0.04\\   
4647 & 6.3   & 0.38 & 2.10  & 0.30 &   1.52 & 0.13 & 48.0 & 0.45 & 0.07 & 0.37 & 0.06 & 0.17 & 0.02\\   
4654 & 7.1   & 0.32 & 2.39  & 0.53 &   1.40 & 0.16 & 48.0 & 0.46 & 0.10 & 0.30 & 0.08 & 0.14 & 0.02\\   
\hline
\end{tabular}

\begin{tabular}{rl}
(1) :& galaxy name (New Galactic Catalogue)\\
(2) :& CO\,(1--0) Intensity $\int T_{\rm MB}dv$ [\rm{K}{ }\rm{km/s}] of the
central position by \citet{Stark86}\\
(3) :& CO\,(2--1) Intensity $\int T_{\rm MB}dv$ [\rm{K}{ }\rm{km/s}] of the
central position observed with KOSMA\\
(6) :& CO\,(3--2) Intensity $\int T_{\rm MB}dv$ [\rm{K}{ }\rm{km/s}] of the
central position observed with KOSMA\\
(8) :& gaussian HPBW of the CO\,(1--0) distribution observed 
by \citet{Ken88}\\ 
(9) :& CO\,(2--1)/(1--0) line ratio corrected for beam filling\\
(11):& CO\,(3--2)/(2--1) line ratio corrected for beam filling\\
(13):& CO\,(3--2)/(1--0) line ratio corrected for beam filling\\
\end{tabular}
}
\end{table*}

\section{Results}

\subsection{The spectra}
From the sample of 20 selected galaxies, we were able to detect 18 in
$^{12}$CO\,(2--1) and 16 in $^{12}$CO\,(3--2) with the KOSMA 3m
telescope. The results of the observations are shown in Fig. 1. The
derived line intensities and intensity ratios are compiled in
Table 2.

NGC\,4303 and NGC\,4321 were mapped at four additional points (five
point cross) to estimate the $^{12}$CO\,(3--2) extend within the larger
$^{12}$CO\,(2--1) beam. The spatial variation of the antenna
temperature is in agreement with the adopted source sizes deduced
below from the published $^{12}$CO\,(1--0) data.

For NGC 4321 higher angular resolution data from the IRAM
30m--telescope is available \citep{sempere97}, which we used to
cross-check the intensity calibration. Fig. 2 shows good agreement
between the KOSMA 3m $^{12}$CO\,(2--1) data set and the convolved IRAM
30m data with regard to both, line shape and calibration.

\begin{figure}[htb]
\resizebox{\hsize}{!}{\includegraphics[angle=-90]{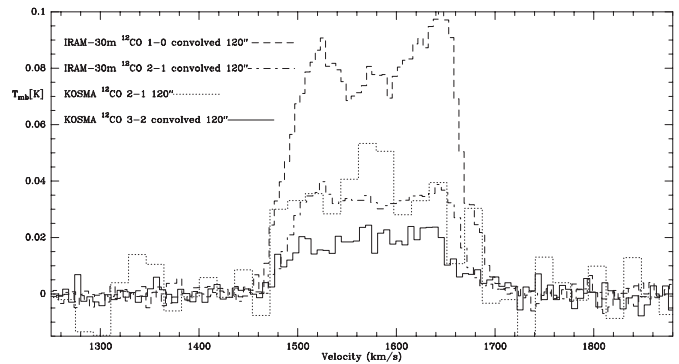}}
\caption{CO in NGC4321: $^{12}$CO\,(1--0) \& $^{12}$CO\,(2--1) 
(convolved to $120''$) spectra using the data observed with the IRAM 30m 
telescope in comparison with the obtained $^{12}$CO\,(2--1) ($120''$) \& 
$^{12}$CO\,(3--2) (convolved to $120''$) KOSMA spectra.}
\end{figure}

\subsection{The Line Ratios}
The main goal of the present work is to derive proper line intensity
ratios. The line widths in CO\,(1--0), (2--1), (3--2) are the same
within the errors (see Fig. 1 and Fig. 2). We thus use the higher S/N
velocity integrated line intensities, rather than peak brightness
temperatures to derive the line intensity ratios.

To correct for the different beam sizes (80\arcsec, 100\arcsec\, and
120\arcsec\, beams) we have to estimate the source extent. Every
galaxy of our survey was also observed in $^{12}$CO\,(1--0) along the
major axis by \citet{Ken88}.  In order to rescale the data to a common
beam size of 120\arcsec\, the radial CO emission profile for each
galaxy can be fitted by a Gaussian function (see Fig. 3).

\begin{figure}[h]
\resizebox{\hsize}{!}{\includegraphics[angle=-90]{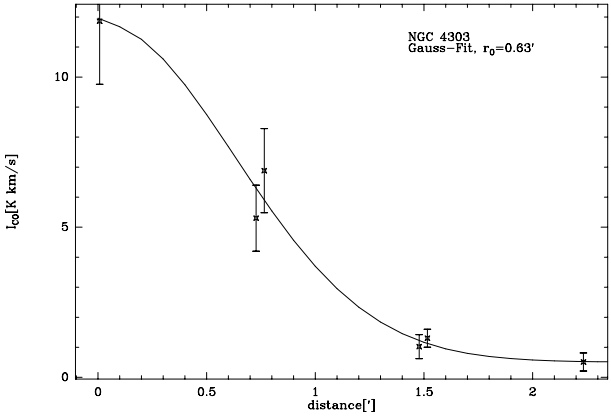}}
\caption{Example for the calculation of the source sizes $\Theta$:
Gaussian fit of the radial CO-distribution \citep{Ken88}
along the major axis of NGC\,4303.}
\end{figure}

Using this we can estimate the extent (Gaussian HPBW: $\Theta$) of the
emitting $^{12}$CO\,(1--0) region.  Assuming that the emitting region
is the same for all three observed transitions it is possible to
rescale each integrated spectrum from its observed beam size $a$ to an
equivalent beam size of 120\arcsec\, (with $a=100''$ for the CO(1--0)
survey by \citet{Stark86} and $a=80''$ for the new KOSMA (3--2)
results):

\begin{equation}
f_{\rm a,120}  = \frac {1+(a/\Theta)^2}{1+(120''/\Theta)^2}.
\end{equation}

We then calculate the corrected line ratios, adopting 
a Gaussian source distribution and a Gaussian profile for a
120\arcsec beam:

\begin{equation}
R_{\rm 2,1}=f^{-1}_{\rm 100,120}\cdot \frac {\int I_{\rm v}(^{12}\rm{CO}(2\to1)) dv}
                               {\int I_{\rm v}(^{12}\rm{CO}(1\to0)) dv},
\end{equation}

\begin{equation}
R_{\rm 3,2}=f_{\rm 80,120}\cdot \frac {\int I_{\rm v}(^{12}\rm{CO}(3\to2)) dv}
                              {\int I_{\rm v}(^{12}\rm{CO}(2\to1)) dv},
\end{equation}

\begin{equation}
R_{\rm 3,1}=\frac {f_{\rm 80,120}}{f_{\rm 100,120}}\cdot \frac {\int I_{\rm v}(^{12}\rm{CO}(3\to2)) dv}
                              {\int I_{\rm v}(^{12}\rm{CO}(1\to0)) dv}.
\end{equation}

The analysis is based on the assumption that the emission from all
three transitions originates from the same region. This seems
justified because on one hand the CO\,(3--2) emission is found
extended beyond the nucleus and also in the discs of spiral galaxies
\citep{Wielbinski02}. On the other hand this is confirmed in our
sample by the detailed study of NGC\,4321 and NGC\,4303, where we
compared R$_{\rm 3,2}$ with the (3--2)/(2--1) ratio when convolving
the CO\,(3--2) map to 120\arcsec\, and found both to be similar within
the errors.

As our beam sizes for the observed transitions differ not that much,
the error due to beam filling is rather small. Even if this assumption
would not apply, the resulting error would nevertheless be relatively
small: The larger scaling factor (i.e.\ the one for the smaller
CO\,(3--2) beam), $f_{\rm 80,120}=(1+(80''/\Theta)^2)/(1+(120''/\Theta)^2$, 
varies only between 0.62 (for the typically largest source sizes
around $80''$ estimated from CO\,(1-0)) and $4/9$ (for compact, point
like emission).

As we lack resolution to solve the problem, we will use the CO\,(1-0)
emission profile to rescale the (3-2) emission.

\begin{figure}[h]
\resizebox{\hsize}{!}{\includegraphics[angle=-90]{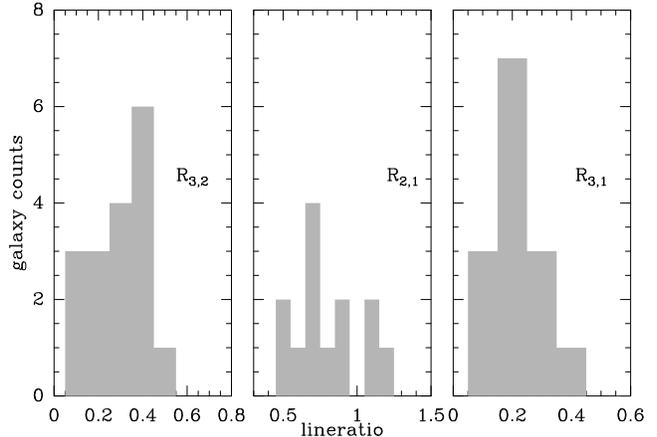}}
\caption{Histogram of the deduced line ratios for a 120\arcsec\,
beam: R$_{\rm 3,2}$,R$_{\rm 2,1}$ and R$_{\rm 3,1}$.}
\end{figure}
               
After correction for beam filling we derive (3--2)/(2--1) line ratios
between 0.1 and 0.45, and (2--1)/(1--0) ratios in the range of 0.45 to
1.2 (excluding the low S/N case of NGC 4192).  The complementary
(3--2)/(1--0) ratios cover a relatively narrow range from 0.11 to
0.35. Table 2 lists the ratios for each source, together with the
error bars estimated from the fit errors of the integrated
intensities; for completeness and easier comparison with the
literature, it also includes the resulting (3--2) versus (1--0) ratio
$R_{\rm 3,1}$. Fig. 4 shows histograms of the observed
distribution of line ratios.

In comparison with this, the CO\,(3--2) survey of 28 nearby galaxies
by \citet{Mauersberger99} covers a range for $R_{\rm 3,1}$ from 0.2 to
0.7, with a tendency for the galaxies without strong star formation
activity towards lower ratios. In contrast, galaxies with star
formation activity such as selected in the survey of 7 nearby
starburst nuclei \citep{Devereux94} or in the study of 8 dwarf
starburst galaxies by \citet{Meier01}, cover higher values between 0.4
and 1.4. In addition, detailed mapping of the CO\,(3--2) emission in a
sample of 12 bright and nearby galaxies by \citet{Dumke01}, shows that
the CO\,(3--2) emission is more compact. The (3--2)/(1--0) ratio drops
from values close to unity in the dense cores of these galaxies, down
to values of around 0.5 with increasing distance from the galactic
nucleus. As mentioned before, all these surveys have been done with
the larger submm telescopes in relatively nearby sources, and thus
have beams that resolve the nucleus/disc structure.  In addition, the
mapping typically is concentrated on the inner regions, where the
CO\,(3--2) emission is still relatively easy to detect. Our, in this
regard unbiased, survey of 20 galaxies in the distant Virgo Cluster
with the comparably large beam of the KOSMA 3m telescope gives a
(3--2)/(1--0) ratio towards the low end or even lower than in the
other studies. This fact supports the concept of a compact, brighter
core region and very weak (if present at all) CO\,(3--2) emission from
the extended discs of galaxies. As at least a substantial fraction of
our Virgo sample is unresolved both in CO\,(1--0) and (3--2), this
ratio will not drop further if averaged over larger areas. We thus
expect the average value of $R_{\rm 3,1}$ of 0.20 to be typical for a
large fraction of galaxies.  A comparable study of a sample of 33
galaxies in the Coma cluster by \citet{Lavezzi99} in the CO\,(2--1)
and (1--0) lines gives a range of (2--1)/(1--0) ratios of 0.5 to 1.3;
this is in good agreement with the range covered by the 16 Virgo
galaxies detected in the present study.

The detected (3--2)/(2--1) ratios have an average value of 0.23, which
is in the order of the line ratio of the core regions of rather
quiescent nearby galaxies as observed by \citet{Mauersberger99} (e.g.
IC\,10, NGC\,891, NGC\,5907).  Observations with the KOSMA 3m
telescope in the galactic ring of the Galaxy (priv. comm. Br\"ull, PhD
work) show an average (3--2)/(2--1) ratio of 0.7\,$\pm$\,0.2\,.

\begin{figure*}[h]
\resizebox{\hsize}{!}{\includegraphics[angle=-90]{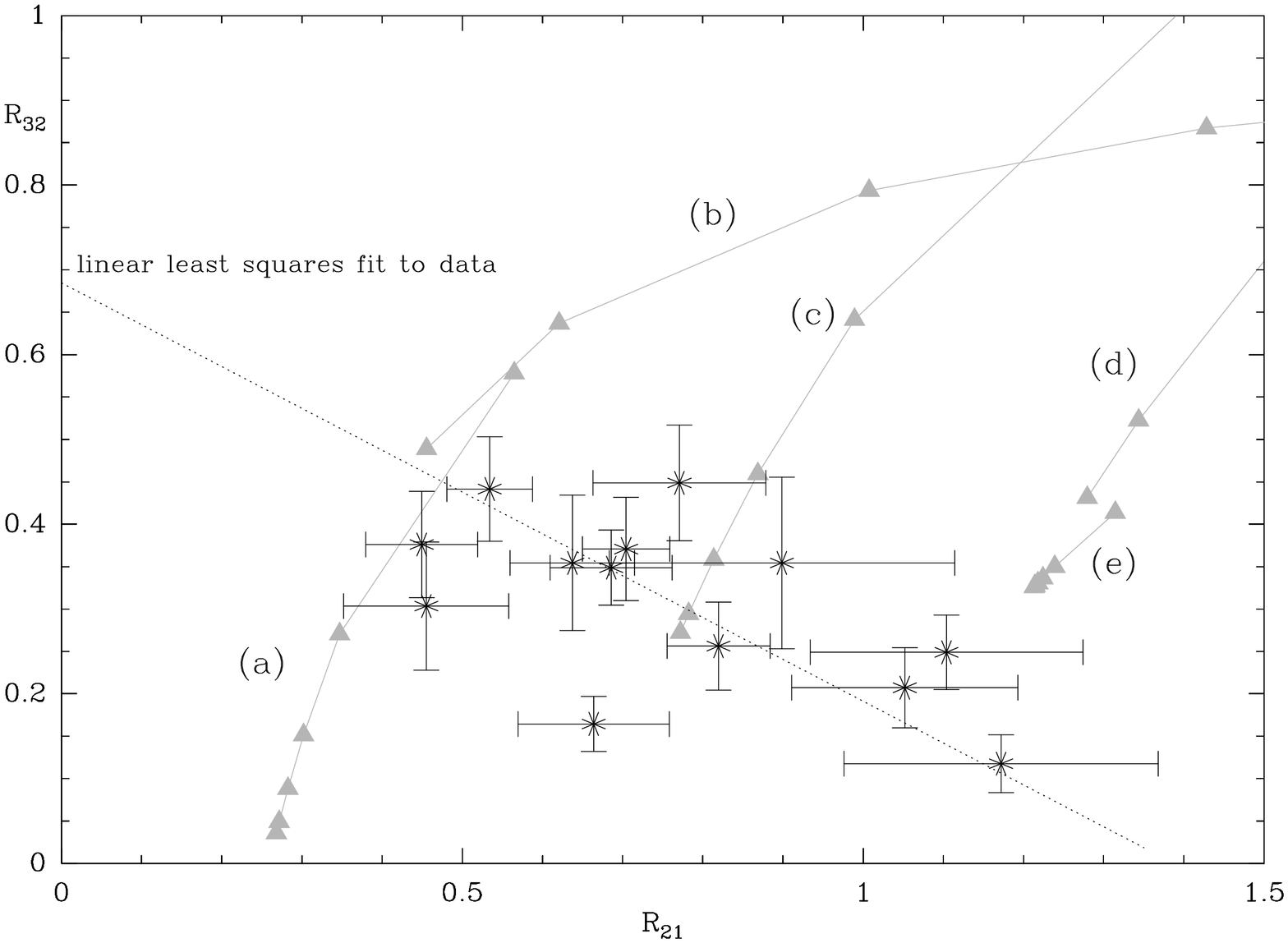}}
\caption{Correlation plot of R$_{\rm 3,2}$, i.e. the CO\,(3--2)/(2--1) 
  line ratios vs. the R$_{\rm 2,1}$ line ratios as observed with the KOSMA
  3m and Bell-Labs 7m telescope for the Virgo sample. The data is
  corrected for main beam efficiency and beam filling. The dotted line
  marks a linear least squares fit to the data. Solid grey lines are
  results computed from a two-component emission model for the galaxies.
  Solid triangles mark the results for a beam filling of the warm dense
  phase of 0.01, 0.02, 0.05, 0.1, 0.2 and 0.5 and typical H$_{\rm 2}$
  densities, CO column densities per velocity interval and kinetic
  temperatures according to the set of model parameters a) to e) from
  Table 3.}
\end{figure*}

\begin{table*}[ht]
\centering{
\footnotesize
\caption{\footnotesize Model parameters for two-component model with 
a warm dense core component and a less dense, cold component. 
Calculated results for the model are plotted in Fig. 5.}
\begin{tabular}{r|lll|lll}
\hline 

Model & cold  &     &                & warm &      &\\
      & T$_{\rm kin}$ & n(H$_{\rm 2}$) & N(CO)/$\Delta$v & T$_{\rm kin}$ & n(H$_{\rm 2}$) & N(CO)/$\Delta$v\\
      & [K]       & cm$^-3$  & cm$^-1$ & [K]       & cm$^-3$  & cm$^-1$\\
\hline
(a)   & 10 & 1000  & 1.5e14 & 50 & 1000  & 1.0e13\\
(b)   & 10 & 1000  & 1.0e14 & 50 & 1.0e4 & 1.0e13\\
(c)   & 10 & 1.0e4 & 1.5e15 & 50 & 1.0e5 & 1.5e15\\
(d)   & 10 & 4.5e4 & 1.5e14 & 50 & 1.0e6 & 1.5e15\\
(e)   & 10 & 4.5e4 & 1.5e14 & 50 & 1.0e4 & 1.0e13\\
\hline
\end{tabular}
}
\end{table*}

\subsection{Modeling of the CO emission line ratios}

Fig. 5 shows $R_{\rm 3,2}$ against $R_{\rm 2,1}$. The resulting
correlation plot does not show any trend for $R_{\rm 3,2}$ growing
with increasing $R_{\rm 2,1}$. This would be expected in a simple
model for the CO emission, where the brightness of the higher
excitation lines would stay high due to the increased temperature or
density of the source region. There seems to be a tendency for an
anti-correlation: Fig. 5 includes the result of a linear least squares
fit analysis, which gives a correlation coefficient of -0.72\,. No
galaxy with $R_{\rm 2,1}$ above unity has $R_{\rm 3,2}$ above
0.3, and all of the galaxies with $R_{\rm 3,2}$ above 0.3 have
$R_{\rm 2,1}$ below 0.8.

Before we can interpret the observed line ratios any further, we check
the data set for any correlation with other parameters of the observed
sample. It is known that the Virgo Cluster galaxies are heavily
HI-deficient in the central region of the cluster \citep{vanGorkom85},
presumably caused by HI ram-pressure stripping due to the interaction
of the galaxies with the cluster environment.  It is important to
check whether this influences the CO emission.  There is no evidence
for a correlation between the CO-emission strength and the distance to
M\,87, which is the kinematic center of the Virgo Cluster (Fig. 6).
This is in agreement with the finding of \citet{Stark86} for the
CO\,(1--0) emission.  The molecular material seems to be affected by
tidal interaction between the cluster
galaxies \citep{Comb88}.

Because of the fact, that the beam covers the central regions of the
galaxies and also partly their discs with various inclination angles
in the sample, one might suspect that optical depth effects and self
absorption of the supposedly brighter emission from the central region
in the disc, might affect the observed line intensities. Regarding to
Fig. 7, no correlation between the observed line ratios and the
inclination angle can be found.

\begin{figure}[htb]
\resizebox{\hsize}{!}{\includegraphics{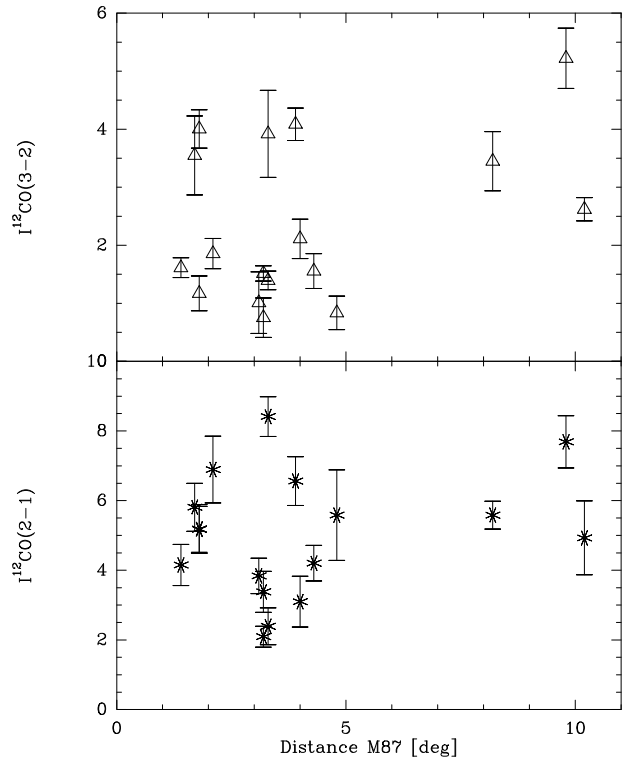}}
\caption{Correlation plot of the integrated line emission I$^{12}$CO\,(3--2)
         (upper panel) and I$^{12}$CO\,(2--1) (lower panel) vs. 
         the angular distance to M87.}
\end{figure}
 
\begin{figure}[htb]
\resizebox{\hsize}{!}{\includegraphics{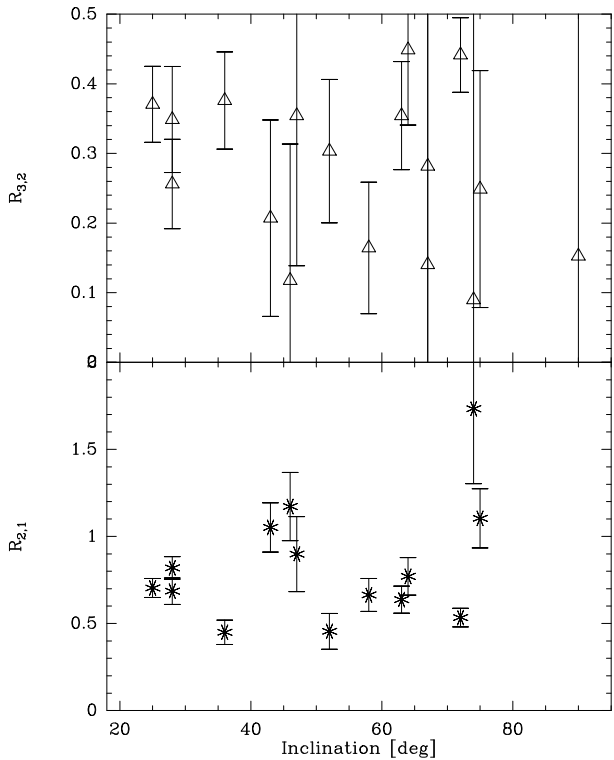}}
\caption{Correlation plot of the  R$_{\rm 2,1}$ (lower panel) and R$_{\rm 3,2}$ 
         (upper panel) line ratios vs. the inclination angle of the
         galaxies.}
\end{figure}

\begin{figure*}[htb]
  \centering 
  \includegraphics[angle=-90,width=13cm]{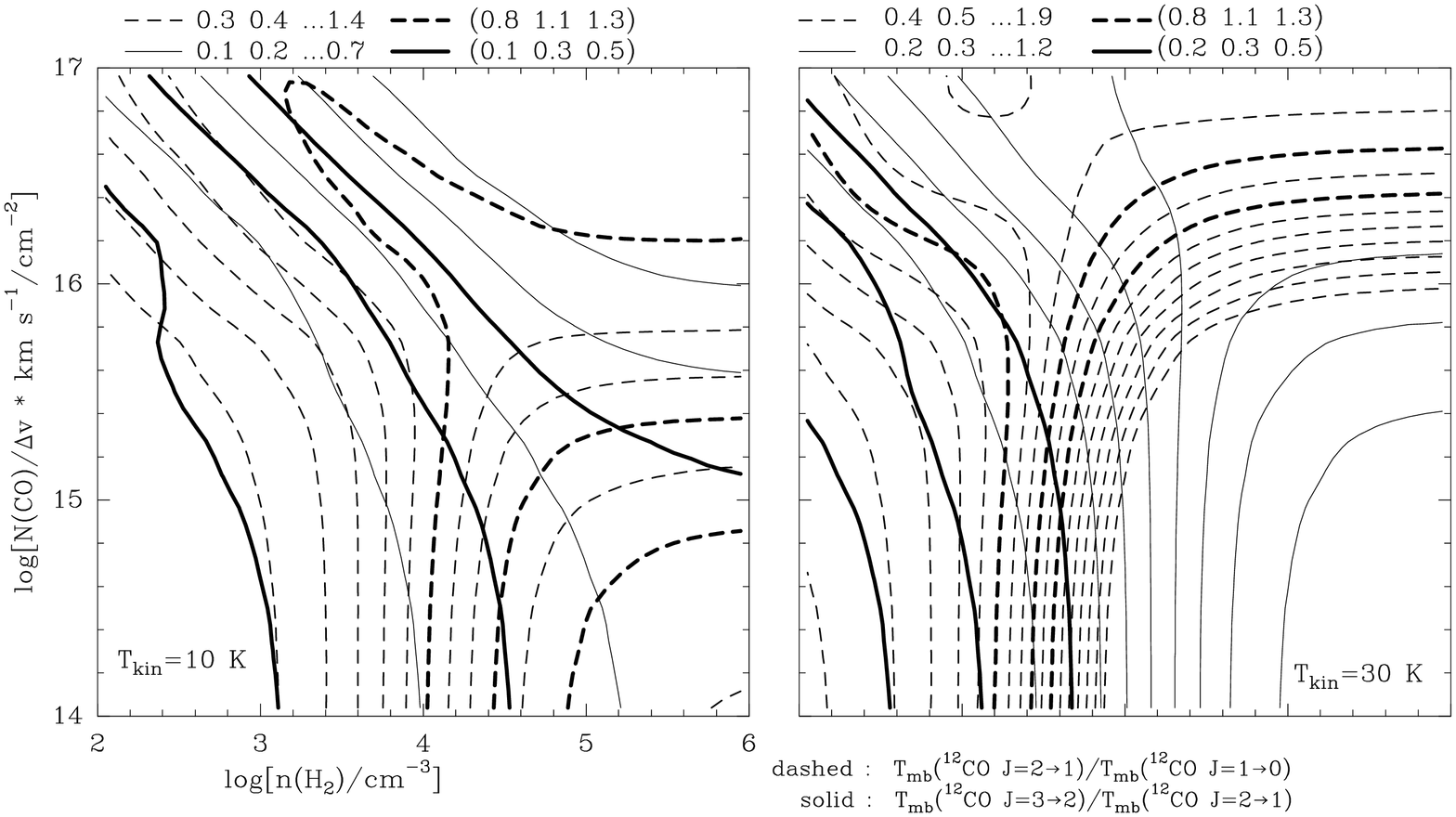}
  \caption{Escape probability-calculations of the CO\,(3--2)/(2--1) and 
           (2--1)/(1--0) ratios at 10\,\rm{K} and at 30\,\rm{K}}
\end{figure*}

\subsubsection{Single component models}

Although the $^{12}$CO emission is almost certainly optically thick,
it is not possible to explain the observed line ratios assuming
LTE-excitation conditions and optically thick $^{12}$CO emission.  We
thus compare the observed ratios with a slightly more sophisticated
excitation model, namely escape probability radiative transfer
calculations for spherical clumps \citep{stutzki85}. In Fig. 8 the
escape probability radiative transfer results are shown for kinetic
temperatures of 10\,\rm{K} and 30\,\rm{K}, and the average observed
line ratios with typical error bars are indicated.

At least the averaged ratios are roughly consistent with the emission
being dominated by very cold gas, 8 to 15\,\rm{K}, at densities of
around $10^4$ cm$^{-3}$ and up to moderate column densities. However,
a detailed comparison of the values for individual sources from Table
2 with the escape probability radiative transfer results shows that
the model fails to explain the two observed ratios within the error
bars in each individual case.

\subsubsection{Two-component model}
 
In a next step we try to model the observed ratios as originating from
a superposition of two components, a warm dense and a cool less dense
gas component, so that

\begin{equation}
T^{\rm b}_{\rm i,j}=\eta[(1-\epsilon)T^{\rm b}_{\rm cold}+\epsilon \cdot T^{\rm b}_{\rm warm}],
\end{equation}
T$^{\rm b}_{\rm i,j}$ the observed brightness temperature of the
observed transition $i\to j$, T$^{\rm b}_{\rm cold}$ the brightness
temperature of the cold phase, $T^{\rm b}_{\rm warm}$ that of the warm
phase, $\epsilon$ the fraction of the warm material, and $\eta$ the
overall beam filling factor of the emission (it is assumed to be the
same for the observed transitions, which cancels out in the calculated
ratios). Obvious choices would be a warm and dense core region filling
only a fraction $\epsilon$ of the beam (and varying in size relative
to the beam from source to source) and a more extended, cold,
low-density component from the galactic disc. The warm, dense
component may also be distributed throughout the disc in the form of
many localized star forming regions. Using escape probability models
we calculated the brightness temperatures for typical H$_{\rm 2}$
densities and CO column densities per velocity interval, varying the
filling factor of the warm phase between 0.01 and 0.5. Selected
results of the models are shown in Fig. 5. The adopted ad hoc chosen
parameters for the models a)$\dots$e) are listed in Table 3. From Fig.
5 it is obvious that it is possible to find solutions for individual
galaxies by interpolating the sample models a)$\dots$e), except for
the galaxies with observed (2--1)/(1--0) ratios above 1 and
(3--2)/(2--1) ratios below 0.3.

The two-component model however fails to explain the apparent
anti-correlation between $R_{\rm 3,2}$ and $R_{\rm 2,1}$, which arises
from the variation of the fraction{ $\epsilon$ alone with otherwise
fixed values for the warm and cold component. The tracks of varying
$\epsilon$ in each model a)$\dots$e) run rather orthogonally to the}
$R_{\rm 3,2}$/$R_{\rm 2,1}$ trend.

\section{Conclusions}
We have investigated a sample of 20 galaxies in the Virgo Cluster,
which have been selected to be the the brightest CO\,(1--0) emitters.
We observed them in the CO\,(3--2) and (2--1) lines using the KOSMA 3m
telescope with its 120\arcsec\,, respectively \, 80\arcsec\,beam. We
detected 18 galaxies in CO\,(2--1) and 16 in CO\,(3--2). This survey
extends the database beyond the handful of quiescent galaxies
investigated in previous surveys.  The relatively large beam includes
the complete CO emission in most cases, so that the correction for
beam filling is relatively straight forward, and we are able to deduce
relatively precise integral line ratios for these sources.

This data set is unique in the sense that it is not biased towards
selecting starburst galaxies or emphasizing the brighter emission from
the core region of galaxies, which has been the tendency in most of the
few previous CO\,(3--2) studies of external galaxies. The CO\,(3--2)
to (1--0) integrated line ratio in our Virgo sample of galaxies covers
a relatively narrow range of values from 0.14 to 0.35, indicating that
most normal galaxies have a line ratio much smaller than that observed
in the dense core regions of nearby galactic centers
or in starburst galaxies.

The observed line ratios $R_{\rm 3,2}$ and $R_{\rm 2,1}$ cannot be
consistently fitted by a single-component radiative transfer model,
thus indicating that at least a two-component model is necessary to
explain the global CO emission properties of the galaxies observed. A
straightforward two-component model, assuming a warm, dense nuclear
and a cold, less dense disc component for all galaxies has a
sufficiently large number of parameters to fit most of the galaxies
individually. It fails, however, to explain the observed trend of line
ratios within the sample galaxies, if the anti-correlation
results only from the variation of the relative filling factor of the
warm gas, with the parameters of the warm, dense and the cold, lower
density component fixed to a set of values typical for all galaxies.

\begin{acknowledgements} 
The KOSMA 3m telescope is operated by the K\"olner Observatorium
f\"ur SubMillimeter Astronomie of the I. Physikalisches Institut,
Universit\"at zu K\"oln in collaboration with the Radioastronomisches 
Institut, Universit\"at Bonn. This project was supported within 
SFB\,301 and SFB\,494 of the DFG and special funding of the Land 
Nordrhein-Westfalen. \\
We thank Garc\'{i}a-Burillo for providing us the IRAM\,30m
data set of NGC\,4321.  We thank C. Kramer, M. Miller, 
A.R.Tieftrunk for help with the observations.\\
\end{acknowledgements} 

\bibliographystyle{aa}
\bibliography{hafokh.bib}

\begin{thebibliography}{18}
\expandafter\ifx\csname natexlab\endcsname\relax\def\natexlab#1{#1}\fi

\bibitem[{{Braine} \& {Combes}(1992)}]{braine92}
{Braine}, J. \& {Combes}, F. 1992, Astronomy and Astrophysics, 264, 433

\bibitem[{{Combes} {et~al.}(1988){Combes}, {Dupraz}, {Casoli}, \&
  {Pagani}}]{Comb88}
{Combes}, F., {Dupraz}, C., {Casoli}, F., \& {Pagani}, L. 1988, Astronomy and
  Astrophysics, 203, L9

\bibitem[{de~Vaucouleurs \& Corwin(1976)}]{deVau76}
de~Vaucouleurs, G. \& Corwin, H.~G. 1976, Second Reference Catalogue of Bright
  Galaxies (Austin: University of Texas Press)

\bibitem[{{Degiacomi} {et~al.}(1995){Degiacomi}, {Schieder}, {Stutzki}, \&
  {Winnewisser}}]{Degiacomi95}
{Degiacomi}, C.~G., {Schieder}, R., {Stutzki}, J., \& {Winnewisser}, G. 1995,
  Optical Engineering, 34, 2701

\bibitem[{{Devereux} {et~al.}(1994){Devereux}, {Taniguchi}, {Sanders}, {Nakai},
  \& {Young}}]{Devereux94}
{Devereux}, N., {Taniguchi}, Y., {Sanders}, D.~B., {Nakai}, N., \& {Young},
  J.~S. 1994, The Astronomical Journal, 107, 2006

\bibitem[{{Dumke} {et~al.}(2001){Dumke}, {Nieten}, {Thuma}, {Wielebinski}, \&
  {Walsh}}]{Dumke01}
{Dumke}, M., {Nieten}, C., {Thuma}, G., {Wielebinski}, R., \& {Walsh}, W. 2001,
  Astronomy and Astrophysics, 373, 853

\bibitem[{Kenney \& Young(1988)}]{Ken88}
Kenney, J.~D. \& Young, J.~S. 1988, The Astrophysical Journal Supplement
  Series, 66, 261

\bibitem[{{Lavezzi} {et~al.}(1999){Lavezzi}, {Dickey}, {Casoli}, \& {Kaz{\`
  e}s}}]{Lavezzi99}
{Lavezzi}, T.~E., {Dickey}, J.~M., {Casoli}, F., \& {Kaz{\` e}s}, I. 1999, The
  Astronomical Journal, 117, 1995

\bibitem[{{Mauersberger} {et~al.}(1999){Mauersberger}, {Henkel}, {Walsh}, \&
  {Schulz}}]{Mauersberger99}
{Mauersberger}, R., {Henkel}, C., {Walsh}, W., \& {Schulz}, A. 1999, Astronomy
  and Astrophysics, 341, 256

\bibitem[{{Meier} {et~al.}(2001){Meier}, {Turner}, {Crosthwaite}, \&
  {Beck}}]{Meier01}
{Meier}, D.~S., {Turner}, J.~L., {Crosthwaite}, L.~P., \& {Beck}, S.~C. 2001,
  The Astronomical Journal, 121, 740

\bibitem[{Sandage \& Tammann(1981)}]{Sandage81}
Sandage, A.~R. \& Tammann, G.~A. 1981, Revised Shapley-Ames Catalog of Bright
  Galaxies (Washington, DC: Carnegie Institutuion of Washington)

\bibitem[{{Sempere} \& {Garc\'{i}a-Burillo}(1997)}]{sempere97}
{Sempere}, M.~J. \& {Garc\'{i}a-Burillo}, S. 1997, Astronomy and Astrophysics,
  325, 769

\bibitem[{Stark {et~al.}(1986)Stark, Knapp, Bally, Wilson, Penzias, \&
  Rowe}]{Stark86}
Stark, A.~A., Knapp, G.~R., Bally, J., {et~al.} 1986, The Astrophysical
  Journal, 310, 660

\bibitem[{Stutzki \& Winnewisser(1985)}]{stutzki85}
Stutzki, J. \& Winnewisser, G. 1985, Astronomy and Astrophysics, 144, 13

\bibitem[{{van Gorkom} \& {Kotanyi}(1985)}]{vanGorkom85}
{van Gorkom}, J. \& {Kotanyi}, C. 1985, in ESO Workshop on the Virgo Cluster,
  61--66

\bibitem[{{Walsh} {et~al.}(2002){Walsh}, {Beck}, {Thuma}, {Weiss},
  {Wielebinski}, \& {Dumke}}]{Wielbinski02}
{Walsh}, W., {Beck}, R., {Thuma}, G., {et~al.} 2002, Astronomy and
  Astrophysics, 388, 7

\bibitem[{{Wielebinski} {et~al.}(1999){Wielebinski}, {Dumke}, \&
  {Nieten}}]{Wielebinski99}
{Wielebinski}, R., {Dumke}, M., \& {Nieten}, C. 1999, Astronomy and
  Astrophysics, 347, 634

\bibitem[{{Young} {et~al.}(1995){Young}, {Xie}, {Tacconi}, {Knezek}, {Viscuso},
  {Tacconi-Garman}, {Scoville}, {Schneider}, {Schloerb}, {Lord}, {Lesser},
  {Kenney}, {Huang}, {Devereux}, {Claussen}, {Case}, {Carpenter}, {Berry}, \&
  {Allen}}]{Young95}
{Young}, J.~S., {Xie}, S., {Tacconi}, L., {et~al.} 1995, The Astrophysical
  Journal Supplement Series, 98, 219

\end{thebibliography}
\end{document}